\documentclass[runningheads]{llncs}
\usepackage{graphicx}
\usepackage{comment}
\usepackage{amsmath,amssymb} 
\usepackage{color}
\usepackage{supertabular}
\usepackage{longtable,booktabs}
 \usepackage{multirow}
\usepackage[perpage,stable]{footmisc} 
\usepackage[width=122mm,left=12mm,paperwidth=146mm,height=193mm,top=12mm,paperheight=217mm]{geometry}
\usepackage[colorlinks,
            linkcolor=blue,
            anchorcolor=blue,
            citecolor=blue]{hyperref}

\begin{document}
\pagestyle{headings}
\mainmatter

\title{	
LIRA: Lifelong Image Restoration from Unknown Blended Distortions} 

\titlerunning{LIRA}
%

\author{Jianzhao Liu\footnotemark[1]\orcidID{0000-0002-5811-0881},
Jianxin Lin\footnotemark[1], Xin Li\orcidID{0000-0002-6352-6523} \and Wei Zhou\orcidID{0000-0003-3641-1429} \and Sen Liu\orcidID{0000-0003-3778-8973} \and Zhibo Chen\footnotemark[2]\orcidID{0000-0002-8525-5066}}
\authorrunning{J. Liu et al.}
\institute{CAS Key Laboratory of Technology in Geo-spatial Information Processing and Application System,\\ University of Science and Technology of China, Hefei 230027, China
\email{\{jianzhao,linjx,lixin666,weichou\}@mail.ustc.edu.cn} \\\email{elsen@iat.ustc.edu.cn}, \quad\email{chenzhibo@ustc.edu.cn}}
\maketitle
\renewcommand{\thefootnote}{\fnsymbol{footnote}} 
\footnotetext[1]{The first two authors contributed equally to this work.} 
\footnotetext[2]{Corresponding author.} 

\begin{abstract}
Most existing image restoration networks are designed in a disposable way and catastrophically forget previously learned distortions when trained on a new distortion removal task. To alleviate this problem, we raise the novel lifelong image restoration problem for blended distortions. 
We first design a base fork-join model in which multiple pre-trained expert models specializing in individual distortion removal task work cooperatively and adaptively to handle blended distortions. When the input is degraded by a new distortion, inspired by adult neurogenesis in human memory system, we develop a neural growing strategy where the previously trained model can incorporate a new expert branch and continually accumulate new knowledge without interfering with learned knowledge. 
Experimental results show that the proposed approach can not only achieve state-of-the-art performance on blended distortions removal tasks in both PSNR/SSIM metrics, but also maintain old expertise while learning new restoration tasks.
\keywords{Image restoration, blended distortions, lifelong learning}
\end{abstract}

\section{Introduction}

Image restoration, which is a highly ill-posed problem, has been studied for a long time due to its high demand in different application scenarios such as surveillance imaging \cite{zhang2010super,svoboda2016cnn} and medical imaging \cite{gondara2016medical,chen2018efficient}. Most existing image restoration methods tend to treat different degradation factors individually and design a dedicated model for each task, which is inefficient and impractical in real world since images are often degraded by various factors with unknown mixture ratios and strengths. Recent studies \cite{yu2018crafting,suganuma2019attention} have been raised for the blended distortion removal task. However, these approaches are designed in a disposable way and lack the ability to maintain previously learned knowledge when dealing with new distortions. 

Unlike deep neural networks that are prone to catastrophic forgetting \cite{2ce896e6d8c74e64888a431e9591fab6,mccloskey1989catastrophic}, human can gradually accumulate knowledge without seriously perturbing past memories \cite{abraham2005memory}  due to special neurophysiological and biological mechanism of learning and memorizing. One explanation about human memory system is stability-plasticity dilemma \cite{mermillod2013stability}, stating that plasticity is required for the integration of new knowledge and stability is required for avoiding perturbation of old memory.

To imitate human memory system and alleviate catastrophic forgetting in image restoration networks, we design an expansible network for lifelong image restoration, in order to accommodate to new blended distortions without forgetting previous restoration tasks. 
Meanwhile, assuming that we cannot access the old training data while training a new task, we leverage a Generative Adversarial Network (GAN) \cite{goodfellow2014generative} to replay learned experience to ensure model stability.

Starting from three typical degradation factors (Gaussian blur, Gaussian noise and JPEG compression), we train a base network in a fork-join manner. In the fork stage, three expert models specializing in specific distortion removal are trained. Then in the join stage, we build the base network by aggregating features from different experts, which helps the three experts work collaboratively to complete a blended distortions removal task. 

When the input is degraded by another new distortion (e.g. haze or darkness), we develop a neural growing strategy where our pre-trained base network can incorporate a new expert branch to form an expanded network and continually accumulate new knowledge. The neural growing strategy is inspired by adult neurogenesis studies \cite{schmidt2004enhanced,kee2007preferential,aimone2011resolving}, which reveals that new neurons are preferentially recruited in response to behavioral novelty and integrated into existing neuronal circuits in selective regions. 

 The base network and the expanded network have shared parameters and constitute an incremental network. The following question is how to adapt the shared parameters to both the old task (dealing with blended distortions formed by the superposition of three degradation factors) and the new task (dealing with blended distortions formed by the superposition of the three degradation factors and another new degradation factor) simultaneously, avoiding the learned knowledge from being overwritten by the new learning knowledge. Studies about Complementary Learning Systems (CLS) \cite{mcclelland1995there} illustrated that recent experience is encoded in the hippocampus and is consolidated in the neocortex through replays of the encoded experience.  Considering that old training data may not be allowed to stored for a long time or shared with others due to storage or privacy issues in real scenarios, we train a GAN to generate pseudo old samples to imitate the memory replay process. The pseudo samples are paired with the corresponding responses of the pre-trained base network, serving as supervision of the old task.  More interestingly, we find that the memory replay strategy can even boost the performance of the old task rather than simply maintain the performance, which means the old task can benefit from incremental learning of new task through our method.

The contributions of this work can be summarized as follows:
\begin{itemize}
    \item We handle the blended distortions in a fork-join manner. Complex tasks are assigned to different expert models. The experimental results show that the expert models dealing with heterogeneous problems can work adaptively in the case of blended distortions. 
    \item We consider lifelong learning in image restoration. We incorporate a new expert branch into the pre-trained base network and leverage a generative adversarial network to replay the learned experience to achieve a trade-off between plasticity and stability. 
\end{itemize}

\section{Related work}
\subsection{Image restoration}
Image restoration tasks, which aim to reconstruct uncorrupted images from corrupted low-quality images, have been widely studied with the rapidly developed CNN-based methods, such as image denoising \cite{zhang2018ffdnet,lefkimmiatis2018universal}, image dehazing \cite{ren2018gated,zhang2018densely} and image deblurring \cite{tao2018scale,nah2017deep}. Liu et al. \cite{liu2019dual} proposed a ``dual residual connection'' which exploits the potential of paired operations. 
Work \cite{liu2019joint} proposed DuRB-M which can handle different degradation factors and demonstrated that multi-task learning of diverse restoration tasks can bring about synergetic
performance improvement. However, images in the real world are usually degraded by complicated blended distortions with unknown mixture ratios. Unlike previous image restoration methods which treat different degradation factors individually and design a dedicated model for each task, some works \cite{yu2018crafting,suganuma2019attention} have been proposed to restore images with blended distortions.  Yu et al. \cite{yu2018crafting} first proposed a reinforcement learning based approach that can progressively restore the corrupted images.
Then Suganuma et al. \cite{suganuma2019attention} presented an operation-wise attention layer, which performs various operations in parallel and weights these operations through an attention mechanism. Although these methods have shown preliminary successes in image restoration from blended distortions, they neglect the problem of how to tackle a novel distortion. It would be important to have an intelligent restorer that could continuously learn to address new distortions, without forgetting how to deal with existing ones. In this paper, we take the first step towards lifelong image restoration from blended distortions, in order to mitigate catastrophic forgetting in image restoration networks.
\subsection{Lifelong learning}
  There have been many works proposed to address the catastrophic forgetting in high-level computer vision tasks such as classification. Regularization-based approaches, such as EWC \cite{kirkpatrick2017overcoming}, HAT \cite{serra2018overcoming} and MAS \cite{aljundi2018memory}, add a regularization term that discourages the alteration to weights important to previous tasks, which effectively prevents old knowledge from being erased or overwritten. Li et al. \cite{li2017learning},  Castro et al. \cite{castro2018end} and Dhar et al. \cite{dhar2019learning} also proposed to employ a distillation loss to encourage the responses of the original and the new network to be similar. 
The rehearsal based  methods \cite{robins1995catastrophic,rebuffi2017icarl,wu2018incremental,lopez2017gradient} refer to relearning representative samples selected from the previous task while learning a new task, which can protect the learned knowledge from disruption by new knowledge. However, due to complexity, privacy and legal issues, previously learned data may not be allowed to be stored for a long period in the real world \cite{wu2018incremental}, which limits the application scenarios of rehearsal based methods. The pseudo-rehearsal based methods  \cite{shin2017continual,kamra2017deep,wu2018memory,wu2018incremental} utilize generative models to generate pseudo-samples for modeling the data distribution of previous tasks and replay learned experiences to consolidate the old task knowledge.  Dynamic architecture methods \cite{rusu2016progressive,yoon2017lifelong} dynamically accommodate new branches or increase the number of trainable parameters to adapt to new tasks. In this paper, we introduce a fork-join model for lifelong image restoration, which has not been explored before. Unlike lifelong learning for classification  aiming at increasing the number of recognizable classes in the network output, our work aims at handling more degradation factors in the input and produce the same clean output. For new distortion coming in, our model only needs to incorporate a new expert branch and continually accumulates new learning knowledge without losing previously learned memory. 

\section{Our Approach}
In the following subsections, we will first introduce the base fork-join network dealing with blended distortions formed by the superposition of \(N\) degradation factors and then demonstrate how to extend the existing base network in the presence of a new degradation factor. Finally, we will describe the training strategy for lifelong image restoration.

\begin{figure*}[ht] 
\centering
\setlength{\belowcaptionskip}{-0.3cm}
\includegraphics[scale=0.38]{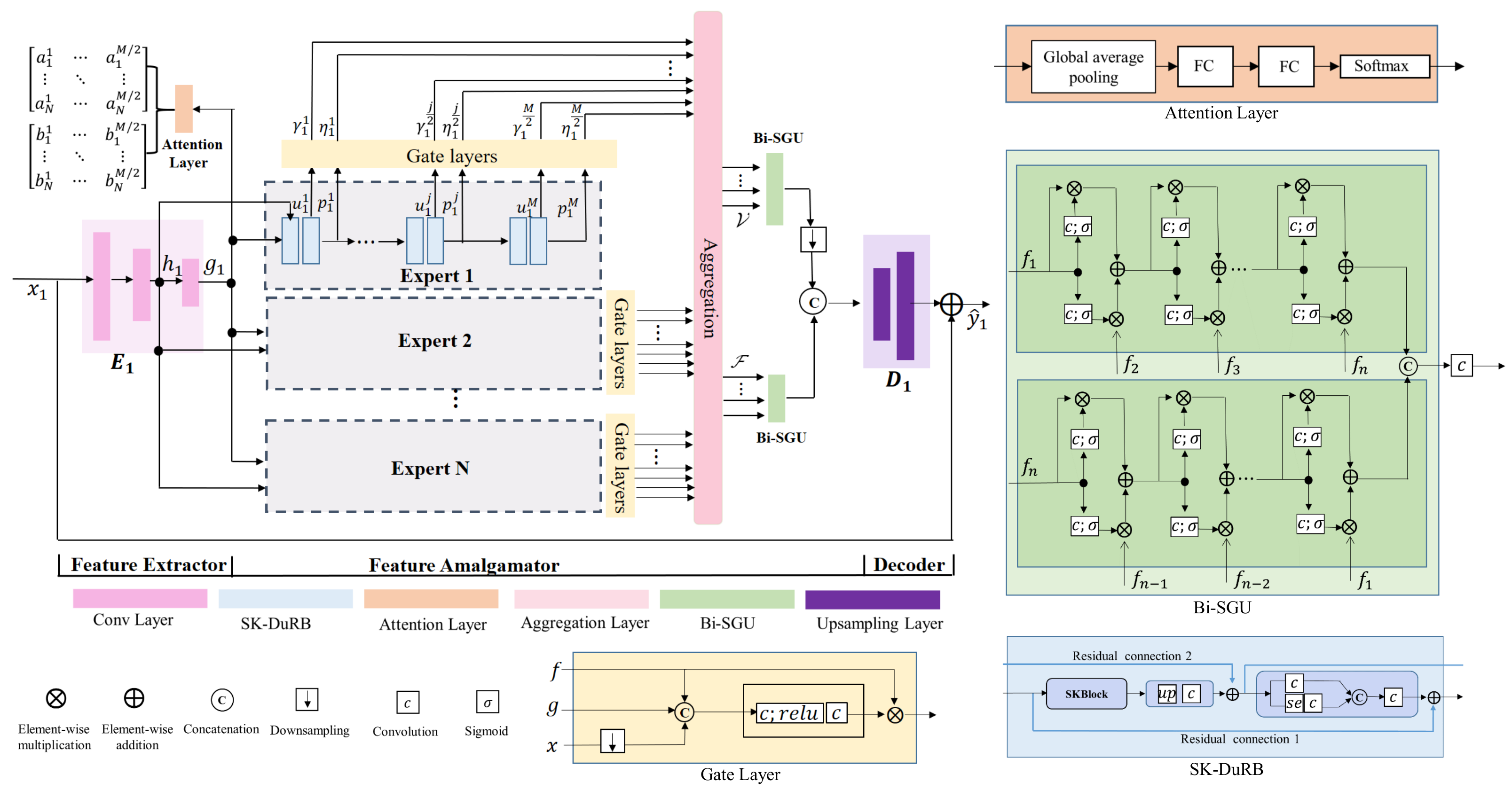} 
\caption{Framework of the base network. It consists of three parts: a feature extractor, a feature amalgamator and a decoder. Expertises of $N$ experts specializing in individual distortion removal are dynamically aggregated for blended distortions removal. $FC$ means Fully Connected layer; $se$ means squeeze-and-excitation block \cite{hu2018squeeze}; $up$ means PixelShuffle  \cite{Shi_2016_CVPR} modules with convolutional operations. In the Gate layer, $f \in  \{\{{s_{i}^{j}}\}_{j=1}^{M/2},\{{q_{i}^{j}}\}_{j=1}^{M/2}\}_{i=1}^{N}$ , $g \in \{ {{h}_{1}},{{g}_{1}}\}$ and $x$ denotes input image. Zooming in for better viewing.}
\label{fig:base_figure} 
\end{figure*}

 \subsection{Base network design}
 The overall architecture of the proposed network is illustrated in Fig. \ref{fig:base_figure}. The entire base network consists of three parts: a feature extractor, a feature amalgamator and a decoder. The feature extractor consists of convolutional layers and produces two-scale feature maps, while the decoder has two upsampling layers that enlarge the spatial size of feature maps. The feature amalgamator consists of \(N\) experts, a set of gate layers, an attention layer, an aggregation layer and two Bidirectional Sequential Gateing Units (Bi-SGUs). Each expert is composed of \(M\) Selective Kernel Dual Residual Blocks (SK-DuRBs). The SK-DuRB is modified from DuRB-M \cite{liu2019joint} by replacing the improved SE-ResNet module with SK-Block \cite{li2019selective} because the DuRB-M has been proved to be robust with different degradation factors and the SK-Block has the capability of adaptively adjusting the receptive fields according to the input.
We first train \(N\) experts each of which specializes in individual distortion removal. Specially, these experts are trained sharing a feature extractor and a decoder. Next, with weight parameters of the $N$ experts loaded, amalgamated features are generated to the decoder for eliminating blended distortions. 

Let's denote the outputs of the \({j}\)-th  SK-DuRB for the \({i}\)-th expert as \(u_{i}^{j}\) and \(p_{i}^{j}\) and take the outputs of two strided convolutional layers in the feature extractor as ${{h}_{1}}$ and ${{g}_{1}}$. ${{h}_{1}}$ and ${{g}_{1}}$ can be seen as the global descriptor for the input images thus we refer to them as global features. Each expert can be depicted as follows:
\begin{equation}
\begin{aligned}
\label{eqn:expert}
\centering
u_{i}^{j},p_{i}^{j}=B_{i}^{j}(u_{i}^{j-1},p_{i}^{j-1}),
\text{1}\le j\le M;1\le i\le N
 \end{aligned}
\end{equation}
where \(u_{i}^{0}={{h}_{1}}\), \(p_{i}^{0}={{g}_{1}}\) and   \(B_{i}^{j}(\cdot )\) refers to the  \({j}\)th  SK-DuRB of the \({i}\)th expert model.  Let \({{s}_{i}}=\{u_{i}^{2j}\}_{j=1}^{M/2}\) and \({{q}_{i}}=\{p_{i}^{2j}\}_{j=1}^{M/2}\) denote features chosen from multiple SK-DuRBs of each expert. As illustrated in \cite{jetley2018learn,schlemper2019attention}, models trained with gates applied to multiple layers implicitly learn to suppress irrelevant regions in an input image while highlighting salient features for a specific task. Therefore, in order to make more use of image information, we employ gate layers (as shown in Fig. \ref{fig:base_figure}) on the intermediate features \({{s}_{i}}\) and \({{q}_{i}}\) for each expert. Meanwhile, gate layers among different experts are expected to learn distortion-specific features. The gate layers produce gated features \(\gamma _{i}^{j}\) and $\eta _{i}^{j}$:
\begin{align}
 \label{eqn:gate1} & \gamma _{i}^{j}=G({{h}_{1}},{{x}_{1}},s_{i}^{j})\odot s_{i}^{j}, \\ 
 \label{eqn:gate2}& \eta _{i}^{j}=G({{g}_{1}},{{x}_{1}},q_{i}^{j})\odot q_{i}^{j},
\end{align}
where $G$ takes as input the intermediate feature ${s_{i}^{j}}$ or ${q_{i}^{j}}$, the global feature ${h}_{1}$ or ${g}_{1}$ and the input image ${x}_{1}$, and generate a pixel-level feature map for ${s_{i}^{j}}$ or ${q_{i}^{j}}$.
We then utilize attention layers to dynamically adjust the mixture ratios of distortion-specific features extracted from different experts thus obtaining two sets of aggregated features $\mathcal{V}$ and $\mathcal{F}$:
\begin{equation}
\label{eqn:att1}
\mathcal{V}=\{{{v}_{j}}\}_{j=1}^{M/2}=\{\sum\limits_{i=1}^{N}{a_{i}^{j}}\gamma _{i}^{j}\}_{j=1}^{M/2}, 
\end{equation}
\begin{equation}
\label{eqn:att2}
   \mathcal{F}=\{{{f}_{j}}\}_{j=1}^{M/2}=\{\sum\limits_{i=1}^{N}{b_{i}^{j}}\eta _{i}^{j}\}_{j=1}^{M/2},
\end{equation}
where \(a_{i}^{j}\) and \(b_{i}^{j}\) are the \((i,j)\) element of ${{(a_{i}^{j})}_{N\times M/2}}$ and ${{(b_{i}^{j})}_{N\times M/2}}$ generated by the attention layers fitted on the global feature \({{g}_{1}}\). $\sum\limits_{i=1}^{N}{a_{i}^{j}}=1$ and $\sum\limits_{i=1}^{N}{b_{i}^{j}}=1$.

$\mathcal{V}$ and $\mathcal{F}$ are composed of features from multiple layers, which contains  multi-level information. In order to more effectively leverage multi-level information, we develop a Bidirectional Sequential Gating Unit (Bi-SGU) inspired by the Sequential Gating Unit \cite{lin2018multi} and BLSTM \cite{zhang2015bidirectional}. Given an activate input \({{f}_{a}}\) and a passive input \({{f}_{b}}\), equation depicting the SGU unit is given as below:
\begin{equation}
    SGU({{\mathbf{f}}_{a}},{{\mathbf{f}}_{b}})=\sigma (conv({{\mathbf{f}}_{a}}))*{{\mathbf{f}}_{a}}+\sigma (conv({{\mathbf{f}}_{a}}))*{{\mathbf{f}}_{b}},
\end{equation}
where \(\sigma (\cdot )\) is a sigmoid activation function and $conv$ denotes convolutional operation. Then the Bi-SGU can be defined as:
\begin{equation}
\begin{split}
Bi\text{-}SGU(\mathcal{V})&=
  conv([ SGU({{v}_{M/2}},SGU({{v}_{M/2-1}},(\cdots ))),\\&SGU({{v}_{1}},SGU({{v}_{2}},(\cdots )))]),
\end{split}
\end{equation}
where \([\cdot ]\) refers to concatenation operation. The final reconstructed image is: 
\begin{equation}
{{\widehat{y}}_{1}}={{D}_{1}}(conv([Bi\text{-}SGU{{(\mathcal{V})}},Bi\text{-}SGU(\mathcal{F})]))+{{x}_{1}},
\end{equation}
where \({{D}_{1}}(\cdot )\) corresponds to the decoder part in Fig. \ref{fig:base_figure}.

In this way, we obtain a base network that can handle $N$ degradation factors. Expertises of different experts can be dynamically aggregated for the blended distortions removal task.

\subsection{Incremental network design}
\begin{figure*}[ht] 
\centering
\setlength{\belowcaptionskip}{-0.3cm}
\includegraphics[scale=0.4]{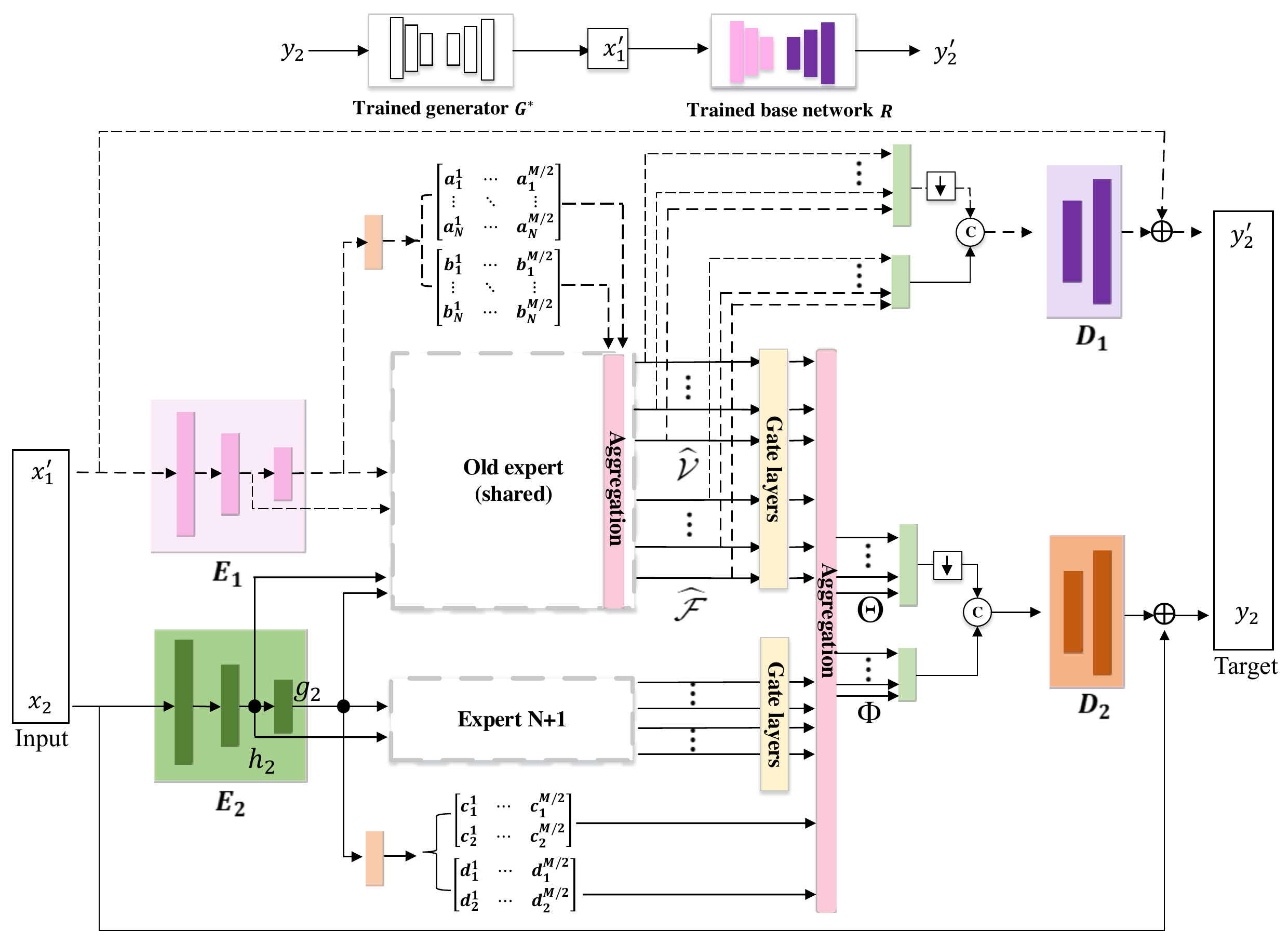}
\caption{Framework of the incremental network. The base network and the expanded network share an old expert and constitute the incremental network. Pseudo-samples ${{x}_{1}'}$ paired with the corresponding responses of the trained base network ${{y}_{2}'}$ are used to optimize the parameters of the base network. The new training samples ${{x}_{2}}$ and ${{y}_{2}}$ are used to train the expanded network.} 
\label{incre_1} 
\end{figure*}

For a new degradation factor, we add a new expert branch and first train it together with the original \(N\) experts whose parameters are frozen, sharing a new feature extractor ${{E}_{2}}$ and a new decoder ${{D}_{2}}$. As  Fig. \ref{incre_1} shows, we assign ${{E}_{2}}$ and ${{D}_{2}}$ to the new task. All parts except Bi-SGUs in the amalgamator of the base network are treated as a whole and named as an old expert afterwards. Now we describe the expanded network in detail below.

For new input image \({{{x}}_{2}}\), the global features extracted from the new feature extractor ${{E}_{2}}$ are denoted as \({{h}_{2}}\) and \({{g}_{2}}\) and the new expert branch can be modeled as:
\begin{equation}
\begin{aligned}
\centering
u_{N+1}^{j},p_{N+1}^{j}=B_{N+1}^{j}(u_{N+1}^{j-1},p_{N+1}^{j-1}),
\text{1}\le j\le M
 \end{aligned}
\end{equation}
where \(u_{N+1}^{0}={{h}_{2}}\) and \(p_{N+1}^{0}={{g}_{2}}\).
Similarly, let ${{\widehat{s}}_{1}}=\widehat{\mathcal{V}}=\{{{\widehat{v}}_{j}}\}_{j=1}^{M/2}$ and ${{\widehat{q}}_{1}}=\widehat{\mathcal{F}}=\{{{\widehat{f}}_{j}}\}_{j=1}^{M/2}$ denote features chosen from the old expert where $\widehat{\mathcal{V}}$ and $\widehat{\mathcal{F}}$ represent aggregated features of the old expert for ${{x}_{2}}$. Let ${{\widehat{s}}_{2}}=\{u_{N+1}^{2j}\}_{j=1}^{M/2}$ and ${{\widehat{q}}_{2}}=\{p_{N+1}^{2j}\}_{j=1}^{M/2}$ denote features chosen from the new expert branch. The aggregated features $\Theta$ and \(\Phi \) for the expanded network are: 
\begin{equation}
\Theta =\{{{\theta }_{j}}\}_{j=1}^{M/2}=\{\sum\limits_{i=1}^{2}{c_{i}^{j}\xi _{i}^{j}}\}_{j=1}^{M/2},
\end{equation}
\begin{equation}
\Phi =\{{{\varphi }_{j}}\}_{j=1}^{M/2}=\{\sum\limits_{i=1}^{2}{d_{i}^{j}\psi _{i}^{j}}\}_{j=1}^{M/2},
\end{equation}
where \(c_{i}^{j}\) and \(d_{i}^{j}\) are the \((i,j)\) element of ${{(c_{i}^{j})}_{2\times M/2}}$ and ${{(d_{i}^{j})}_{2\times M/2}}$ generated by the attention layers fitted on the global feature \({{g}_{2}}\). $\sum\limits_{i=1}^{2}{c_{i}^{j}}=1$ and $\sum\limits_{i=1}^{2}{d_{i}^{j}}=1$. \(\xi _{i}^{j}\) and \(\psi _{i}^{j}\) are given by the formulas below:
\begin{align}
  & \xi _{i}^{j}=G({{h}_{2}},{{x}_{2}},\widehat{s}_{i}^{j})\odot \widehat{s}_{i}^{j}, \\ 
 & \varphi _{i}^{j}=G({{g}_{2}},{{x}_{2}},\widehat{q}_{i}^{j})\odot \widehat{q}_{i}^{j}.
\end{align}
The final reconstructed image for the new task is given as follow:
\begin{equation}
{{\widehat{y}}_{2}}= {{D}_{2}}(conv([Bi\text{-}SGU{{(\Theta )}},Bi\text{-}SGU(\Phi )]))+{{x}_{2}}.
\end{equation}

In this way, the new expert branch can be assimilated into the base network, forming an expanded network to handle new blended distortions.  We finally get an incremental network consisting of the base network and the expanded network. It should be noticed that old expert’s parameters are shared by the base network and the expanded network and are dynamically optimized for both old and new tasks.

\subsection{Training strategy for lifelong image restoration}\label{sec:train_strategy}  
     Our training strategy is inspired by the Complementary Learning System(CLS) theory \cite{mcclelland1995there},  which illustrates that experience encoded in the hippocampus is consolidated in the neocortex through memory replay. 
     We leverage a generative adversarial network to model the data distribution of old training samples, playing the role of memory replay. As shown at the top of Fig. \ref{incre_1}, we first learn a mapping function \(G\) \cite{zhu2017unpaired} aiming to convert clean images ${y}_{1}$ to blended distortion images ${x}_{1}$ of the old task.
     We denote the data distribution as \({{p}_{data}}({{x}_{1}})\) and \({{p}_{data}}({{y}_{1}})\). The adversarial loss and the cycle consistency loss can be depicted as follows:
\begin{equation}
\begin{split}
   {{\mathcal{L}}_{GAN}}&={{\mathbb{E}}_{{{x}_{1}}\sim {{p}_{data}}({{x}_{1}})}}[\log {{D}_{{{X}_{1}}}}({{x}_{1}})]\\&+{{\mathbb{E}}_{{{y}_{1}}\sim {{p}_{data}}({{y}_{1}})}}[\log (1-{{D}_{{{X}_{1}}}}(G({{y}_{1}})))],
\end{split}
\end{equation}
\begin{equation}
  {{\mathcal{L}}_{cyc}}={{\mathbb{E}}_{{{y}_{1}}\sim {{p}_{data}}({{y}_{1}})}}[{{\left\| R(G({{y}_{1}}))-{{y}_{1}} \right\|}_{1}}],
\end{equation}
where \(G\) learns to generate blended distorted images, \({{D}_{{{X}_{1}}}}\) is the discriminator network leaning to distinguish real distorted images and generated ones. \(R\) is the previously trained base network on the old task. The full objective is:
\begin{equation}
\label{GAN total loss}
\begin{split}
 \mathcal{L}={{\mathcal{L}}_{GAN}}+{{\lambda }_{1}} {{\mathcal{L}}_{cyc}},
 \end{split}
\end{equation}
where \({{\lambda }_{1}}\) controls the relative importance of the two objectives. 

After obtaining the trained generator \({{G}^{*}}\), we can generate pseudo old training samples \({{x}_{1}'}={{G}^{*}}({{y}_{2}})\) and \({{y}_{2}'}=R({{G}^{*}}({{y}_{2}}))\) using training samples \(\{{{x}_{2}}_{_{i}},{{y}_{2}}_{_{i}}\}_{i=1}^{T}\) of the new task where $T$ is the number of the new task training samples.  In the following step, we train the incremental network with new training samples and pseudo old training samples. Let's denote the base network as \(\widetilde{R}\) with parameters \({{\theta }_{Base}}=\{{{\theta }_{{S}}},{{\theta }_{{{P}_{1}}}}\}\) and denote the expanded network as $H$ with parameters  \({{\theta }_{E\text{xpanded}}}=\{{{\theta }_{{S}}},{{\theta }_{{{P}_{2}}}}\}\).  ${{\theta }_{{S}}}$ refers to the shared parameters (the parameters of the old expert) between the base network and the expanded network, while ${{\theta }_{{{P}_{1}}}}$ and ${{\theta }_{{{P}_{2}}}}$ denote specific parameters of the base network and the expanded network respectively.

First, the base network \(\widetilde{R}\) is initialized with the previously trained base network $R$ and the expanded network is initialized with the trained ($N+1$)th expert. Then we jointly optimize \({{\theta }_{Base}}\) and \({{\theta }_{E\text{xpanded}}}\) to minimize loss for both two tasks:
\begin{equation}
\centering
\label{incremental_loss}
{{\mathcal{L}}_{incremental}}={{\mathcal{L}}_{new}}({{y}_{2}},{{\widehat{y}}_{2}})+{{\lambda }_{2}} {{\mathcal{L}}_{old}}({{y}_{2}'}, {{\widehat{{y}}_{2}'}}), 
\end{equation}
where \({{\widehat{y}}_{2}}=H({{x}_{2}};{{\theta }_{E\text{xpanded}}})\), ${{\widehat{{y}}_{2}'}}=\widetilde{R}({{x}_{1}'};{{\theta }_{Base}})$ and ${{\lambda }_{2}}$ is a loss balance weight to balance the old task performance and the new task performance. ${\mathcal{L}}_{new}$ and ${\mathcal{L}}_{old}$ are both \({{l}_{1}}\) loss functions.

Through this training strategy, the incremental network can continually accumulate new knowledge while consolidating previously learned knowledge through memory replay, without accessing the old training samples. 

\section{Experiments}
\subsection{Datasets} 
Following the experimental settings of works \cite{yu2018crafting,suganuma2019attention}, we use the DIV2K dataset \cite{agustsson2017ntire} which contains 800 images (750 images for training and 50 images for testing). The images are corrupted by a sequence of Gaussian blur, Gaussian noise and JPEG compression with random levels and are then cropped into \(64\times 64\) sub-images, resulting in total 230,139 training images and 3,584 testing images.  The images for training experts that specialize in individual distortion removal, are generated in the same way and individual degradation factor is randomly added. In order to verify the effectiveness of our incremental network, we add another two degradation factors (haze and darkness) respectively to the above three degradation factors with random levels and form another two training sets with 230,139 images and two testing sets with 1,827 images. Specially, to further verify the generalization ability of our base network on images from different datasets and in different resolutions, we randomly add blended distortions on five public benchmark datasets: BSDS100 \cite{arbelaez2010contour}, MANGA109 \cite{matsui2017sketch}, SET5 \cite{bevilacqua2012low}, SET14 \cite{zeyde2010single} and URBAN100 \cite{huang2015single} and randomly crop the images into \(64\times 64\), \(128\times 128\) and \(256\times 256\) respectively, generating the additional testing sets. 
 
 The standard deviations of Gaussian blur and Gaussian noise are uniformly distributed in [0, 5] and [0, 50] respectively. 
 We use the ``imwrite'' function in MATLAB to apply JPEG compression, where the parameter ``quality'' within the range of [0,100] controls the compression quality.
 We randomly choose the ``quality'' from the range of [10,100].   For haze degradation, we randomly choose the scattering coefficient within [0.05, 0.25]. For darkness degradation, we use the ``imadjust'' function in MATLAB to randomly apply a gamma adjustment within the range of [1.5, 2.5]. The images can be classified into three groups based on the applied degradation levels: mild, moderate and severe. To test the generalization ability of the network, the training is performed on the moderate group and testing is performed on all three groups (DIV2K dataset). BSDS100, MANGA109, SET5, SET14 and URBAN100 are added with moderate-level distortions.

\subsection{Implementation details}
 \label{sec:details}
In our basic network, we set the number of experts $N$ to 3 and the number of SK-DuRBs $M$ to 6. For model training, we use Adam solver \cite{kingma2014adam} with parameters ${{\beta }_{1}}=0.9$ and ${{\beta }_{2}}=0.999$ . We adopt cosine annealing strategy to adjust the initial learning rate $1{{e}^{-4}}$. For the adversarial generator used to generate pseudo-samples, we adopt the same architecture as the adversarial generator $G$ with 9 residual blocks in \cite{zhu2017unpaired} and replace the adversarial generator $F$ in  \cite{zhu2017unpaired} with our trained base network. We set \({{\lambda }_{1}}\) in Eqn. \ref{GAN total loss} to 10.0 and  \({{\lambda }_{2}}\) in Eqn. \ref{incremental_loss} to 0.2. Specially, in order to discourage the alteration to the parameters closely related to the old task, we lower down the learning rate of \({{\theta }_{Base}}\) and set $lr({{\theta }_{{{P}_{2}}}})=\mu$ and $lr({{\theta }_{S}},{{\theta }_{{{P}_{1}}}})=\rho \mu$, where \(lr(\cdot )\) represents the learning rate and $\rho$ is a small number within the range of $(0,1)$.  We set $\mu$, and $\rho$ to $1{{e}^{-4}}$ and $1/1000$ respectively. We take 5 epochs to warm up the network (freeze the parameters of the old expert and only train the expanded network) and then jointly optimize all parameters.

\subsection{Performance of base network}
We compare our base network with RL-Restore \cite{yu2018crafting} and Operation-wise Attention Network (OWAN) \cite{suganuma2019attention} under PSNR and SSIM metrics. RL-Restore and OWAN are state-of-the-art models that are specially designed for blended distortions. Table \ref{DIV2K comparison} shows the PSNR and SSIM values of the three methods at different degradation levels on DIV2K test sets. It can be seen that our method outperforms the compared methods at all degradation levels in both the PSNR and SSIM metrics.
\begin{table}[h]
\caption{Quantitative results of RL-Restore, OWAN and our base network on DIV2K test sets. The best results are highlighted in bold.}
\centering
\setlength{\abovecaptionskip}{-0.4cm}
\setlength{\belowcaptionskip}{-1cm}
\label{DIV2K comparison}
\scalebox{1}{
\begin{tabular}{|c|l|l|l|l|l|l|}
\hline
\multirow{2}{*}{Method} & \multicolumn{2}{c|}{Mild(unseen)} & \multicolumn{2}{c|}{Moderate} & \multicolumn{2}{c|}{Severe(unseen)} \\ \cline{2-7} 
 & \multicolumn{1}{c|}{PSNR} & \multicolumn{1}{c|}{SSIM} & \multicolumn{1}{c|}{PSNR} & \multicolumn{1}{c|}{SSIM} & \multicolumn{1}{c|}{PSNR} & \multicolumn{1}{c|}{SSIM} \\ \hline
RL-Restore & 28.04 & 0.6498 & 26.45 & 0.5587 & 25.20 & 0.4777 \\
OWAN & 28.33 & 0.7455 & 27.07 & 0.6787 & 25.88 & 0.6167 \\
Ours & \textbf{28.61} & \textbf{0.7496} & \textbf{27.24} & \textbf{0.6832} & \textbf{25.93} & \textbf{0.6219} \\ \hline
\end{tabular}}
\end{table}

\begin{figure*}[ht!]
\setlength{\abovecaptionskip}{0cm}
\setlength{\belowcaptionskip}{0cm}
\includegraphics[scale=0.38]{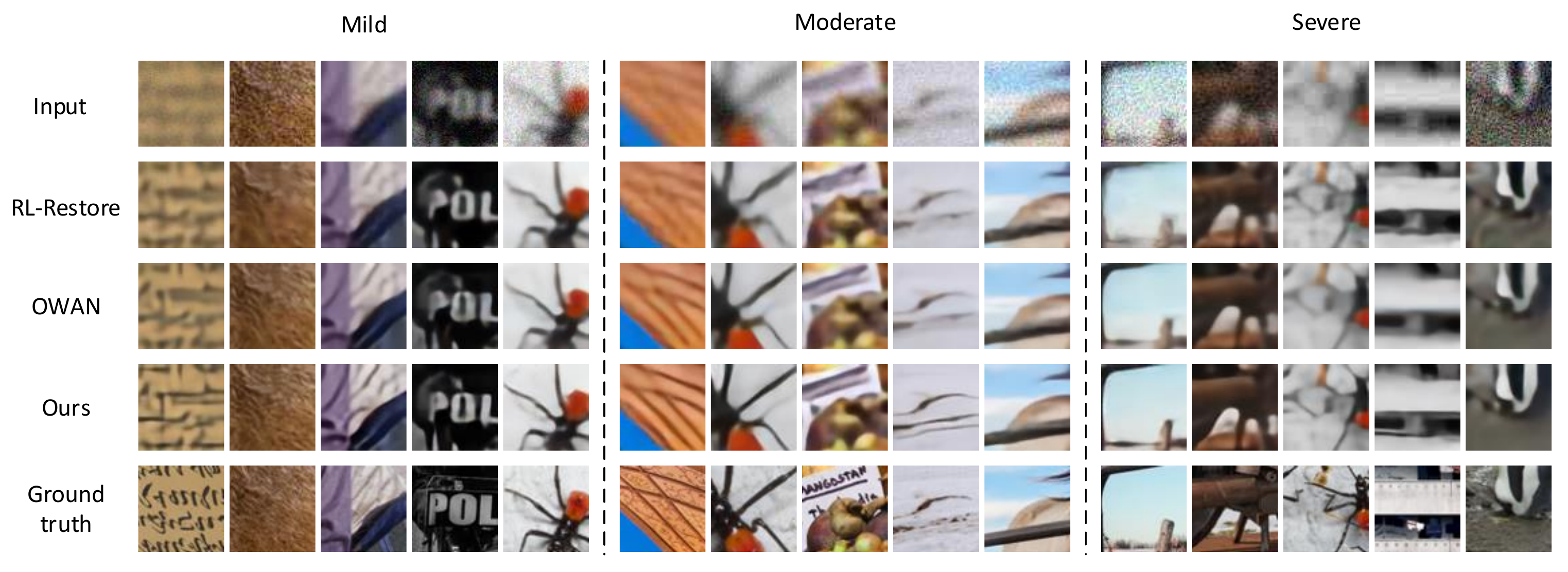} 
\caption{Qualitative comparisons with RL-Restore and OWAN on DIV2K test sets.} 
\label{visual comparison} 
\end{figure*}

\begin{table}[t!]
\caption{Quantitative results of RL-Restore, OWAN and our base network  on BSDS100, MANGA109, SET5, SET14 and URBAN100. The best results are highlighted in bold.}
\centering
\label{classical comparison}
\scalebox{0.85}{
\begin{tabular}{|l|l|l|l|l|l|l|l|l|l|l|l|}
\hline
\multirow{2}{*}{Method} & \multicolumn{1}{c|}{\multirow{2}{*}{\begin{tabular}[c]{@{}c@{}}Image\\ size\end{tabular}}} & \multicolumn{2}{c|}{BSDS100} & \multicolumn{2}{c|}{MANGA109} & \multicolumn{2}{c|}{SET5} & \multicolumn{2}{c|}{SET14} & \multicolumn{2}{c|}{URBAN100} \\ \cline{3-12} 
 & \multicolumn{1}{c|}{} & PSNR & SSIM & PSNR & SSIM & PSNR & SSIM & PSNR & SSIM & PSNR & SSIM \\ \hline
RL-Restore & \multirow{3}{*}{$64\times64$} & 25.57 & 0.6074 & 25.70 & 0.7292 & 25.46 & 0.6549 & 24.68 & 0.6198 & 25.01 & 0.6354 \\
OWAN &  & 25.99 & 0.6261 & 26.68 & 0.7648 & 26.14 & 0.6874 & 25.39 & 0.6567 & 25.64 & 0.6721 \\
Ours &  & \textbf{26.21} & \textbf{0.6298} & \textbf{27.69} & \textbf{0.7751} & \textbf{26.49} & \textbf{0.6955} & \textbf{25.78} & \textbf{0.6652} & \textbf{25.97} & \textbf{0.6783} \\ \hline
RL-Restore & \multicolumn{1}{c|}{\multirow{3}{*}{$128\times128$}} & 24.60 & 0.6108 & 24.58 & 0.7263 & 25.98 & 0.7005 & 24.30 & 0.6725 & 23.87 & 0.6227 \\
OWAN & \multicolumn{1}{c|}{} & 23.70 & 0.6127 & 21.66 & 0.7143 & 23.52 & 0.7005 & 22.53 & 0.6650 & 22.65 & 0.6341 \\
Ours & \multicolumn{1}{c|}{} & \textbf{25.24} & \textbf{0.6377} & \textbf{26.13} & \textbf{0.7741} & \textbf{27.58} & \textbf{0.7637} & \textbf{25.10} & \textbf{0.7079} & \textbf{24.72} & \textbf{0.6641} \\ \hline
RL-Restore & \multirow{3}{*}{$256\times256$} & 23.56 & 0.5724 & 23.15 & 0.6908 & 28.86 & 0.7881 & 24.58 & 0.6974 & 22.54 & 0.6077 \\
OWAN &  & 22.50 & 0.5745 & 19.44 & 0.6549 & 20.66 & 0.6726 & 21.73 & 0.6686 & 21.37 & 0.6136 \\
Ours &  & \textbf{24.10} & \textbf{0.6038} & \textbf{24.45} & \textbf{0.7428} & \textbf{28.86} & \textbf{0.8098} & \textbf{25.33} & \textbf{0.7328} & \textbf{23.39} & \textbf{0.6564} \\ \hline
\end{tabular}
}
\end{table}

 We also show restored images obtained by different methods along with their input images and ground truths in  Fig. \ref{visual comparison}. It can be seen that the visual results are consistent with quantitative results, where our base network can achieve the best visual quality and tends to retain more image detail information. Furthermore, we carry out extensive experiments on five public benchmark datasets: BSDS100, MANGA109, SET5, SET14 and URBAN10 and the quantitative results are shown in Table \ref{classical comparison}. We can see that our network has the best generalization ability on different datasets and has better robustness in terms of resolution variations.


 \subsection{Ablation studies}
In this section, we conduct several ablation studies to investigate the effectiveness of pre-trained experts, the SK-DuRB, the gate layer, the attention layer and the Bi-SGU, which are all the basic components in our base network. So total six variant networks are designed for comparison: (a) To validate the effectiveness of the pre-trained experts, we train the base network without the initialization of experts. (b) To validate the effectiveness of the SK-DuRB, we replace the SK-DuRB with the DuRB-M \cite{liu2019joint}. (c) To verify the effectiveness of the gate layer, we  remove the gate layers applied to $s_{i}^{j}$ in Eqn. \ref{eqn:gate1}. Then we further remove all gate layers applied to $q_{i}^{j}$ in Eqn. \ref{eqn:gate2}. (d) To verify the effect of the attention layer, we set all attention weights equal to 1. (e) To verify the effect of the Bi-SGU, we replace the Bi-SGU with the concatenation operation followed by a convolutional layer. We train these networks using the same strategy as aforementioned in Section \ref{sec:details}. Table \ref{ablation} shows that the original base network achieves the best performance among these variants at all degradation levels in terms of PSNR and SSIM (on DIV2K testing set). 
 \begin{table}[th!]
\caption{Ablation study of our base network.}
\centering
\label{ablation}
\scalebox{1}{
\begin{tabular}{|l|c|c|c|c|c|c|}
\hline
\multicolumn{1}{|c|}{\multirow{2}{*}{Model}} & \multicolumn{2}{c|}{Mild(unseen)} & \multicolumn{2}{c|}{Moderate} & \multicolumn{2}{c|}{Severe(unseen)} \\ \cline{2-7} 
\multicolumn{1}{|c|}{} & PSNR & SSIM & PSNR & SSIM & PSNR & SSIM \\ \hline
w/o Experts & 27.28 & 0.7009 & 25.95 & 0.6370 & 24.72 & 0.5832 \\ \hline
w/o SK-Blocks & 28.14 & 0.7232 & 26.76 & 0.6588 & 25.41 & 0.6010 \\ \hline
w/o Half of Gate layers & 28.51 & 0.7437 & 27.14 & 0.6783 & 25.85 & 0.6191 \\ \hline
w/o All Gate layers & 27.96 & 0.7186 & 26.64 & 0.6560 & 25.38 & 0.6018 \\ \hline
w/o Attention layer & 28.10 & 0.7236 & 26.74 & 0.6591 & 25.39 & 0.6027 \\ \hline
w/o Bi-SGUs & 28.52 & 0.7435 & 27.18 & 0.6789 & 25.91 & 0.6209 \\ \hline
Ours & \textbf{28.61} & \textbf{0.7496} & \textbf{27.24} & \textbf{0.6832} & \textbf{25.93} & \textbf{0.6219} \\ \hline
\end{tabular}}
\end{table}




\subsection{Performance of incremental network}
We have three settings for training the incremental network:


\noindent\textbf{Setting A}. Only real new training samples \(\{{{x}_{2}}_{_{i}},{{y}_{2}}_{_{i}}\}_{i=1}^{T}\) are used to train the expanded network $H$. We discourage alteration to the old expert's parameters by reducing the learning rate. 

\noindent\textbf{Setting B (Joint learning)}. Real new training samples \(\{{{x}_{2}}_{_{i}},{{y}_{2}}_{_{i}}\}_{i=1}^{T}\) and real old training samples \(\{{{x}_{1}}_{_{i}},{{y}_{1}}_{_{i}}\}_{i=1}^{K}\) are used to optimize the parameters of the expanded network $H$ and the base network $\widetilde{R}$ respectively, with the learning rate of the base network lowered. $K$ is the number of the old task training samples.

\noindent\textbf{Setting C (LIRA)}. As illustrated in Section \ref{sec:train_strategy}, real new training samples \(\{{{x}_{2}}_{_{i}},{{y}_{2}}_{_{i}}\}_{i=1}^{T}\) and pseudo old training samples $\{{{x}_{{{1}_{i}}}'},{{y}_{{{2}_{i}}}'}\}_{i=1}^{T}$ are used to optimize the parameters of the expanded network $H$ and the base network $\widetilde{R}$ respectively, with the learning rate of the base network lowered. 
\begin{figure}[t!] 
\centering
\setlength{\abovecaptionskip}{0.2cm}
\setlength{\belowcaptionskip}{-0.5cm}
\includegraphics[scale=0.48]{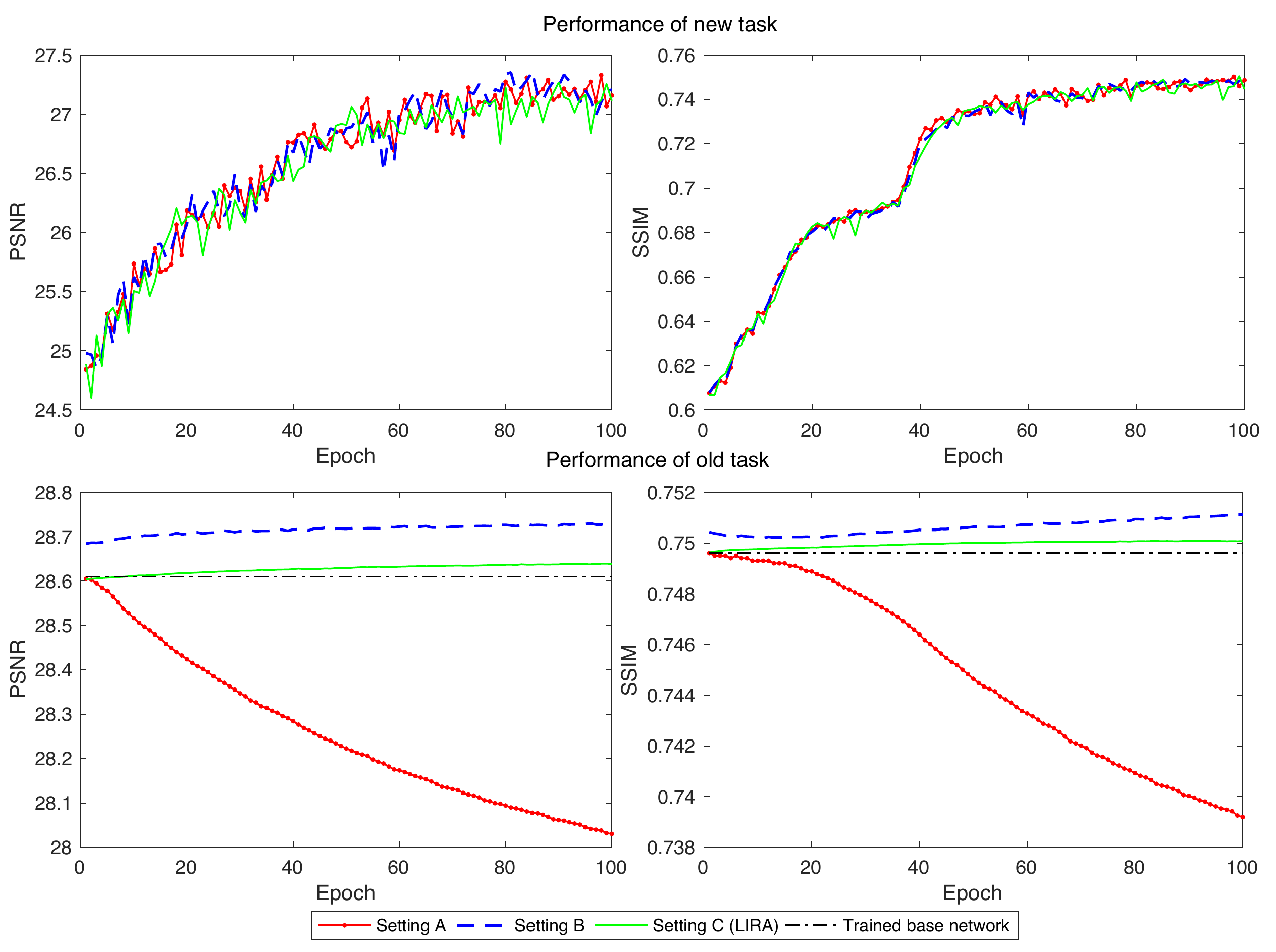} 
\caption{PSNR and SSIM curves with respect to training epochs of 3 different settings. } 
\label{fig:curve_5set} 
\end{figure}

We present the training curves of the three settings on mild DIV2K test group in Fig. \ref{fig:curve_5set}. 
It should be mentioned that we initialize the expanded network $H$ with trained parameters obtained in the warm-up step under the three settings, thereby eliminating random initialization interference under the three settings. 
We can see that the performance on the old task continually drops under setting A.
In contrast, Setting B and LIRA can
even boost the performance of the old task rather than simply maintain the performance, since the new task can provide complementary knowledge for the old task by updating the shared parameters. 
It should be noted that  setting B can be seen as an upper bound on the performance of LIRA since an adversarial generator cannot fully resemble the data distribution of the real old samples. 

In addition to the above three settings, we also compare LIRA with the following baselines: \textbf{(a) Elastic Weight Consolidation (EWC)} \cite{kirkpatrick2017overcoming}: Add a regularization term to discourage the alteration to weights important to the old task. We implement EWC on the old expert's parameters to find which weights are important for the old task when training the new task. \textbf{(b) Progressive Neural Networks (PNN)} \cite{rusu2016progressive}: Allocate a new column whose weights are randomly initialized for each new task while freezing the weights of previous columns. Transfer of old knowledge is enabled via lateral connections to features of previously learned columns. We implement PNN by freezing the old expert's weights and randomly initializing the weights of the new expert branch. \textbf{(c) Learning without Forgetting (LWF)} \cite{li2017learning}: Responses to the new task from the previously trained network are used to optimize old task's performance while new training data is used to optimize new task's performance. We implement LWF by using new training samples \(\{{{x}_{2}}_{_{i}},{{y}_{2}}_{_{i}}\}_{i=1}^{T}\) and responses to the new task from trained base network \(\{{{x}_{2}}_{_{i}},R({{x}_{2}}_{_{i}})\}_{i=1}^{T}\) to train the expanded network $H$ and the base network $\widetilde{R}$ respectively, where $R$ denotes previously trained base network.   \textbf{(d) Fine-tuning}: Modify the parameters of an existing network to train a new task. To implement this method, we train the expanded network $H$ using only real new training samples \(\{{{x}_{2}}_{_{i}},{{y}_{2}}_{_{i}}\}_{i=1}^{T}\). The difference between Fine-tuning and Setting A is that we don't specially lower down the learning rate of the old expert and all parameters are optimized with the same learning rate.  \textbf{(e) OWAN} \cite{suganuma2019attention}: We merge old and new training samples together to train OWAN, whose parameters are initialized with trained weights for old task. 

The comparison results of old and task performance regarding haze degradation factor are shown in Table \ref{lifelong_compare}.  We can see that EWC cannot effectively prevent performance degradation of old task.  Although PNN can guarantee the old task performance, it limits the new task performance because the old expert's parameters are frozen. LWF can achieve relatively good performance on old task but it still causes the old task performance to slightly drop. For Fine-tuning, the descending rate of the old task performance is faster compared with Setting A, because there is no constraints to the old expert's learning rate and the shared parameters will be more adaptive to the new task. Moreover, since the data distribution of the two training sets is different, it is difficult to find a local minimum that is most suitable for both old and new tasks for OWAN. Therefore, both the old and the new task performances of OWAN are worse than our model. Compared with the above methods, LIRA can maintain old expertise while accumulating new knowledge without accessing old training samples. 
\begin{table}[h]
\setlength{\abovecaptionskip}{-0.3cm}
\setlength{\belowcaptionskip}{0.2cm}
\caption{Quantitative results of old and new tasks regarding haze degradation factor. The top three results are highlighted in red, blue and cyan colors respectively in each column.}
\label{lifelong_compare}
 \scalebox{0.9}{
 \centering
\begin{tabular}{|c|l|l|l|l|l|l|}
\hline
\multirow{2}{*}{Method} & \multicolumn{2}{c|}{Mild(unseen)} & \multicolumn{2}{c|}{Moderate} & \multicolumn{2}{c|}{Severe(unseen)} \\ \cline{2-7} 
 & \multicolumn{1}{c|}{\begin{tabular}[c]{@{}c@{}}old\\ PSNR/SSIM\end{tabular}} & \multicolumn{1}{c|}{\begin{tabular}[c]{@{}c@{}}new\\ PSNR/SSIM\end{tabular}} & \multicolumn{1}{c|}{\begin{tabular}[c]{@{}c@{}}old\\ PSNR/SSIM\end{tabular}} & \multicolumn{1}{c|}{\begin{tabular}[c]{@{}c@{}}new\\ PSNR/SSIM\end{tabular}} & \multicolumn{1}{c|}{\begin{tabular}[c]{@{}c@{}}old\\ PSNR/SSIM\end{tabular}} & \multicolumn{1}{c|}{\begin{tabular}[c]{@{}c@{}}new\\ PSNR/SSIM\end{tabular}} \\ \hline
EWC & 22.05/0.4327 & 27.16/0.7485 & 20.86/0.3449 & 25.79/0.6750 & 19.61/0.2644 & 23.92/0.5795 \\
PNN & \textcolor{cyan}{28.61/0.7496} & 24.78/0.6119 & \textcolor{blue}{27.24/0.6832} & 24.21/0.6037 & \textcolor{blue}{25.93/0.6219} & 22.68/0.4706 \\
LWF & 28.56/0.7496 & 27.23/0.7493 & 27.13/0.6804 & 25.97/0.6767 & 25.81/0.6191 & 24.30/0.5887 \\
Setting A & 28.03/0.7394 & 27.33/0.7502 & 26.50/0.6633 & \textcolor{blue}{26.06/0.6781} & 25.13/0.5950 & \textcolor{cyan}{24.31/0.5899} \\
Fine-tuning & 20.25/0.3959 & \textcolor{red}{27.51/0.7566} & 19.13/0.3127 & 25.90/0.6783 & 17.88/0.2311 & 24.26/0.5907 \\
OWAN & 27.57/0.7459 & 27.18/0.7539 & 26.30/0.6785 & 25.71/0.6731 & 25.02/0.6097 & 24.28/0.5891 \\
Joint learning & \textcolor{red}{28.73/0.7511} & \textcolor{blue}{27.39/0.7527} &\textcolor{red}{ 27.25/0.6837} & \textcolor{red}{26.11/0.6793} &\textcolor{cyan}{25.90/0.6228} & \textcolor{red}{24.36/0.5917} \\
LIRA & \textcolor{blue}{28.63/0.7501} & \textcolor{cyan}{27.37/0.7509}& \textcolor{cyan}{27.24/0.6830} & \textcolor{cyan}{26.05/0.6772} &\textcolor{red} {25.93/0.6223} & \textcolor{blue}{24.36/0.5901} \\ \hline
\end{tabular}}
\end{table}

\section{Conclusions}
In this paper, we take the first step toward lifelong image restoration in the presence of blended distortions. To alleviate the catastrophic forgetting in image restoration networks, our model LIRA is designed by imitating human memory system.  Meanwhile, we leverage a GAN to replay learned experience to ensure model stability. As a result, LIRA can continually accumulate new knowledge and retain previous capabilities without accessing old training samples. Extensive experiment results show that LIRA achieves state-of-the-art performance on both ordinary image restoration tasks and lifelong image restoration tasks.  

\section*{Acknowledgements}
This work was supported in part by NSFC under Grant U1908209, 61632001 and the National Key Research and Development Program of China 2018AAA0101400.


\clearpage
%
%
\bibliographystyle{splncs04}
\bibliography{egbib}

\section*{Appendix}

In this supplementary material, we  provide additional results to complement the paper: (1) We provide the network parameters and configuration of our base network, our expanded network, the generator and the discriminator in Section \ref{sec:network architecture}. (2) We show comparisons with the state-of-the-art algorithms in terms of complexity in Section \ref{Sec:Comparisons in terms of complexity}. 
 (3) We show qualitative results of pseudo-samples generated by the generator in Section \ref{subsec:Qualitative results of pseudo samples}.

\subsection*{Network Architectures}
\label{sec:network architecture} 
Our incremental network consists of the base network and the expanded network. We also employ a generator that generates pseudo samples to consolidate the knowledge of the old task. Table \ref{tab:base network architecture} and Table \ref{tab:expanded network} list the detailed configuration and parameters of our base network and our expanded network respectively. Table \ref{tab: generator} shows the architecture of the generator and Table \ref{tab: discriminator} shows the architecture of the discriminator.

\setlength{\tabcolsep}{0.3pt}
\setlength\LTleft{-0.1in}
{
\small
\begin{longtable}{|l|l|l|l|l|l|l|l|l|}
\caption{\textbf{Detailed architecture of our base network.} It consists of a feature extractor, a feature amalgamator and a decoder. $[\cdot ]$ refers to concatenation operation. The feature amalgamator consists of 3 experts, a set of gate layers, an attention layer, an aggregation layer and two Bidirectional Sequential Gateing Units (Bi-SGUs). In the decoder, we use PixelShuffle  \cite{Shi_2016_CVPR} module as the upsampling layer to upsample the feature maps by $2\times$. The final reconstructed image is the summation of the output of the $Decoder1$ (out\_up1-2) and the input image.}
\label{tab:base network architecture} \\
\hline
\multicolumn{2}{|l|}{Base network}                                                                                                           & Input                                                                                                                                                                                                                                                                                                                                                                                                                                                                                                        & Output                                                                                                                                                       & \begin{tabular}[c]{@{}l@{}}Kernel\\  size\end{tabular} & \begin{tabular}[c]{@{}l@{}}In-\\put\\ cha-\\nnels\end{tabular} & \begin{tabular}[c]{@{}l@{}}Out-\\put\\ cha-\\nnels\end{tabular} & \begin{tabular}[c]{@{}l@{}}Str-\\ide\end{tabular} & \begin{tabular}[c]{@{}l@{}}Output \\ size\end{tabular} \\
\hline
\end{longtable}}

\setlength{\tabcolsep}{0.3pt}
\setlength\LTleft{-0.2in}
{
\small
\begin{longtable}{|l|l|l|l|l|l|l|l|l|}
\caption{\textbf{Detailed architecture of our expanded network.} We keep the intact architecture of the base network and all parts except Bi-SGUs in the amalgamator of the base network are treated as a whole and named as an old expert. The final reconstructed image is the summation of the output of the $Decoder2$ (out\_up2-2) and the input image.}
\label{tab:expanded network}\\
\hline
\multicolumn{2}{|l|}{Expanded network}                                                                                                        & Input                                                                                                                                                                                                                                            & Output                                                                                                                                                                                     & \begin{tabular}[c]{@{}l@{}}Kernel\\ size\end{tabular} & \begin{tabular}[c]{@{}l@{}}In-\\put\\ cha-\\nnels\end{tabular} & \begin{tabular}[c]{@{}l@{}}Out-\\put\\ cha-\\nnels\end{tabular}              & \begin{tabular}[c]{@{}l@{}}Str-\\ide\end{tabular} & \begin{tabular}[c]{@{}l@{}}Output\\ size\end{tabular}                                      \\
\hline
\end{longtable}}

\begin{table}[]
\centering
\caption{\textbf{Detailed architecture of the generator.} We use the transposed convolution to upsample the feature maps by $2\times$.}
%
 \\ \hline
conv1       & $7\times7$                                                     & 3                                                        & 64                                                        & 1      & $64\times64$                                                 \\ \hline
conv2       & $3\times3$                                                    & 64                                                       & 128                                                       & 2      & $32\times32$                                                 \\ \hline
conv3       & $3\times3$                                                      & 128                                                      & 256                                                       & 2      & $16\times16$                                                 \\ \hline
$9\times$ResBlock  &$3\times3$                                                      & 256                                                      & 256                                                       & 1      & $16\times16$                                                 \\ \hline
upsampling1 &$3\times3$                                                      & 256                                                      & 128                                                       & 1/2    & $32\times32$                                                 \\ \hline
upsampling2 &$3\times3$                                                      & 128                                                      & 64                                                        & 1/2    & $64\times64$                                                 \\ \hline
conv4       &$7\times7$                                                      & 64                                                       & 3                                                         & 1      & $64\times64$                                                 \\ \hline
\end{tabular}
\label{tab: generator}
\end{table}

\begin{table}[]
\centering
\caption{\textbf{Detailed architecture of the discriminator.} The input is a $64\times64$ RGB image. The output is a single scalar.}
\begin{tabular}{|l|l|l|l|l|l|}
\hline
Layer                                                     & \begin{tabular}[c]{@{}l@{}}Kernel\\ size\end{tabular} & \begin{tabular}[c]{@{}l@{}}Input\\ channels\end{tabular} & \begin{tabular}[c]{@{}l@{}}Output\\ channels\end{tabular} & Stride & \begin{tabular}[c]{@{}l@{}}Output\\ size\end{tabular} \\ \hline
conv1                                                     & $4\times4$                                                   & 3                                                        & 64                                                        & 2      & $32\times32$                                                 \\ \hline
conv2                                                     & $4\times4$                                                   & 64                                                       & 128                                                       & 2      & $16\times16$                                                 \\ \hline
conv3                                                     & $4\times4$                                                   & 128                                                      & 256                                                       & 2      & $8\times8$                                                   \\ \hline
conv4                                                     & $4\times4$                                                   & 256                                                      & 512                                                       & 1      & $8\times8$                                                   \\ \hline
conv5                                                     & $4\times4$                                                   & 512                                                      & 1                                                         & 1      & $8\times8$                                                   \\ \hline
\begin{tabular}[c]{@{}l@{}}Average\\ pooling\end{tabular} & $8\times8$                                                   & 1                                                        & 1                                                         & $8\times8$    & $1\times1$                                                  \\ \hline
\end{tabular}
\label{tab: discriminator}
\end{table}

\newpage
\subsection*{Additional Experimental Results}
\subsubsection*{Comparisons in terms of complexity}
\label{Sec:Comparisons in terms of complexity}
The complexity of RL-Restore \cite{yu2018crafting}, OWAN \cite{suganuma2019attention}, our base network and our expanded network can be found in Table \ref{tab:complexity comparison}. We compute the complexity of these methods on a \(63\times 63\) image. It can be seen that the complexity of our base network is lower than that of OWAN. In addition, the complexity of the expanded network increases by just $33.4\%$ over that of the base network. In contrast, if we train two individual networks for two tasks respectively, the overall complexity of the networks is twice as that of the base network.

\begin{table*}[ht]
\centering
\caption{Comparisons with DnCNN, RL-Restore and OWAN in terms of complexity.}
\label{tab:complexity comparison}
\begin{tabular}{|l|l|l|l|l|l|}
\hline
Methods  &  RL-Restore \cite{yu2018crafting} & OWAN \cite{suganuma2019attention}  & Our base network & Our expanded network \\ \hline
FLOPs(G) &  0.474      & 1.439 & 1.259            & 1.679                \\ \hline
\end{tabular}
\end{table*}

\subsubsection*{Qualitative results of pseudo samples}
\label{subsec:Qualitative results of pseudo samples}
\begin{figure*}[ht!] 
\centering
\includegraphics[scale=0.64]{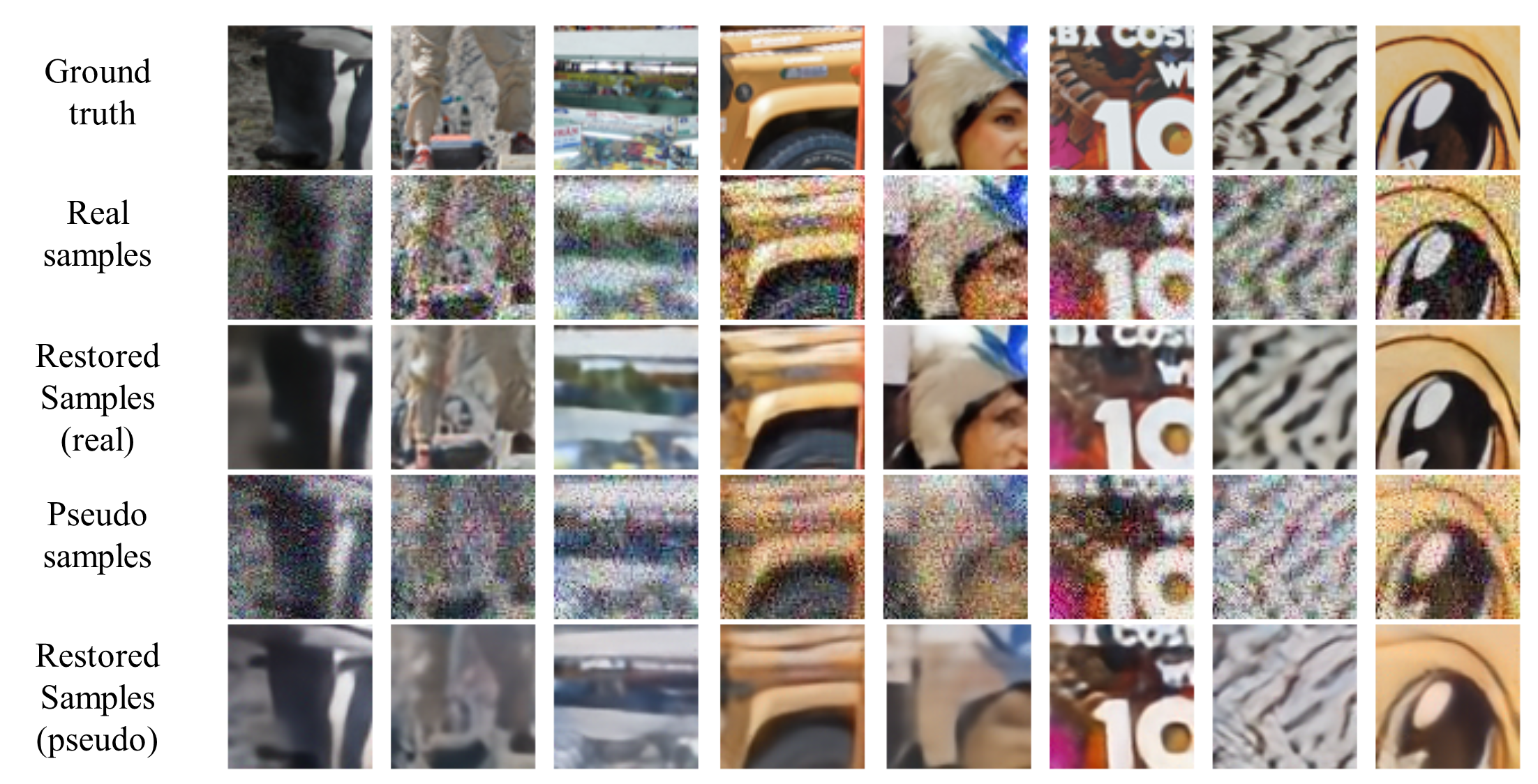} 
\caption{Qualitative results of real and pseudo samples with corresponding restored samples generated by our trained base model.} 
\label{fig:pesudo samples } 
\end{figure*}

For training our incremental network, pseudo samples generated by the generator are paired  with  the corresponding responses of our trained base network, serving as supervision of the old task. We show examples of real and pseudo samples with corresponding restored samples generated by our trained base model in Fig. \ref{fig:pesudo samples }. We can see that the generator can resemble the data distribution of the real samples to some extent, playing the role of memory replay to consolidate the knowledge of the old task.

\end{document}